\documentclass{article}
\usepackage{spconf,amsmath,graphicx}
\usepackage{amssymb}
\usepackage{multirow}
\usepackage{booktabs}
\usepackage{ctable}
\usepackage[utf8x]{inputenc}

\title{Towards Natural and Controllable Cross-Lingual Voice Conversion Based on Neural TTS Model and Phonetic Posteriorgram}
\name{Shengkui Zhao, Hao Wang, Trung Hieu Nguyen, Bin Ma}
\address{Speech Lab, Alibaba Group}
\begin{document}
%
\maketitle
\begin{abstract}
Cross-lingual voice conversion (VC) is an important and challenging problem due to significant mismatches of the phonetic set and the speech prosody
of different languages.  In this paper, we build upon the neural text-to-speech (TTS) model, i.e., FastSpeech, and LPCNet neural vocoder to design a new cross-lingual VC framework named FastSpeech-VC.  We address the mismatches of the phonetic set and the speech prosody by applying Phonetic PosteriorGrams (PPGs), which have been proved to bridge across speaker and language boundaries. Moreover, we add
normalized logarithm-scale fundamental frequency (Log-F0) to further compensate for the prosodic mismatches and significantly improve naturalness. Our experiments on English and Mandarin languages demonstrate that with only mono-lingual
corpus, the proposed FastSpeech-VC can achieve high quality converted speech with  mean opinion score (MOS) close to the professional records while maintaining good speaker similarity. Compared to the baselines using Tacotron2 and Transformer TTS models, the FastSpeech-VC can achieve controllable converted speech rate and much faster inference speed. More importantly, the FastSpeech-VC can easily be adapted to a speaker with limited training utterances.

\end{abstract}
\section{Introduction}

The primary goal of cross-lingual voice conversion (VC) is to convert one's utterances uttered in one language to sound as if they are produced by another speaker who speaks a different language.  The cross-lingual VC has been applied in computer-assisted pronunciation training \cite{Felps2009, Probst2002, Ding2019} and cross-lingual text-to-speech (TTS) systems \cite{sun2016tts, zhang2019, Zhao2020}.

Comparing to the mono-lingual VC where the source speaker and the target speaker are from the same language, the cross-lingual VC is more challenging due to significant mismatches between phonetic sets and speech prosody of different languages. Moreover, the parallel training data, where the same text spoken by both source and target speakers is very difficult to obtain if possible. Most of the VC techniques \cite{Mashimo2002, qian2011, wang2015, 2019skzhao} that rely on speaker-dependent features have difficulties bridging across language boundaries.
\begin{figure}[t]
  \centering
  \includegraphics[width=\linewidth]{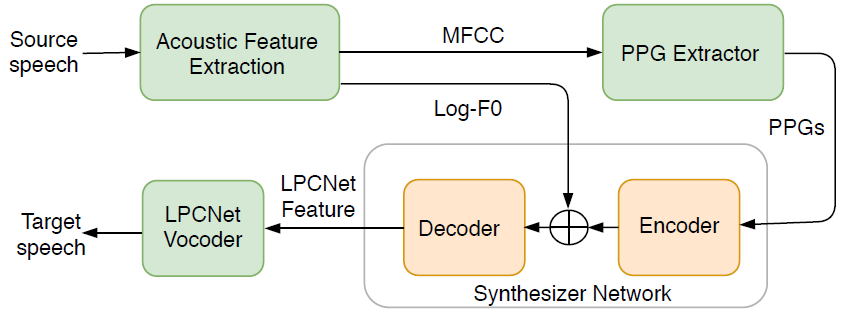}
  \caption{Our proposed FastSpeech-VC framework for cross-lingual VC task.}
  \label{fig:vc_framework}
\end{figure}
Recently, Phonetic PosteriorGram (PPG) was introduced to address the phonetic and  prosody mismatches for cross-lingual VC task \cite{sun2016, zhou2019}. PPGs represent posterior probabilities of phonetic classes on frame level, which are independent of speakers and languages. The works \cite{sun2016, zhou2019} tried to map PPGs to Mel-Cepstral Coefficient (MCC) sequences using a target speaker's DBLSTM model, where the output MCC sequence is synthesized to the target speaker's voice by STRAIGHT \cite{sun2016} or WORLD \cite{Morise16} vocoders. Both the DBLSTM models and the conventional vocoders tend to produce characteristic artifacts, resulting in lower audio quality. Moreover, we found that PPGs alone do not preserve prosody
well and the logarithm-scale fundamental frequency (Log-F0) is important to the naturalness of converted speech.

In this paper, we build upon the recent studies of neural FastSpeech text-to-speech (TTS) \cite{Ren2019} model  and neural LPCNet vocoder \cite{Valin2019} to design a new cross-lingual (VC) framework. We address the phonetic and prosodic mismatches by applying Phonetic PosteriorGrams (PPGs) and normalized logarithm-scale fundamental frequency (Log-F0). The proposed cross-lingual VC framework is named as FastSpeech-VC. Different from the works \cite{sun2016, zhou2019}, FastSpeech-VC extends the original FastSpeech TTS model for VC task. FastSpeech-VC not only dramatically improves the inference speed with parallel non-autoregressive sequence generation, but also achieves controllable speech rate. Moreover, the LPCNet vocoder is  a trainable neural vocoder, which produces high-quality reconstructed speech. Our experiments on English and Mandarin languages demonstrate that with only mono-lingual
corpus, the proposed FastSpeech-VC can generate high quality converted speech with  mean opinion score (MOS) close to professional records while archiving good speaker similarity. Compared to the baselines using Tacotron2 \cite{Jonathan2018} and Transformer \cite{Naihan2018} TTS models, the FastSpeech-VC can further achieve controllable converted speech rate and much faster inference speed. More importantly, the FastSpeech-VC model can be easily adapted to a speaker with limited training utterances.

\section{The Cross-Lingual FastSpeech-VC Framework}
Our proposed FastSpeech-VC framework is shown in Fig \ref{fig:vc_framework}. The framework is composed of three major components: a PPG extractor, a synthesizer network, and a LPCNet vocoder. The framework first uses the PPG extractor to extract a PPG sequence from speech frames. The PPG sequence is then synthesized to the LPCNet acoustic feature sequence using the synthesizer network that bases on neural FastSpeech model. The LPCNet acoustic feature sequence is finally converted to speech frames by the LPCNet vocoder.

\subsection{The PPG extractor}
The PPG extractor transforms speaker-dependent acoustic features to Phonetic PosteriorGrams, that is, the frame based posterior probabilities of phonetic classes (phonemes or triphones/senones). PPGs can retain acoustic information while excluding speaker identities, therefore suitable for cross-lingual VC by bridging the acoustic information across speakers and languages.

In this work, we use a DNN-based acoustic model from an ASR system to perform as the PPG extractor.  The PPG extractor is trained to classify frame-based MFCCs to the corresponding senone classes by minimizing cross-entropy loss. In our experiments, the senone labels were first obtained from force-alignments of a well-trained GMM-HMM system. Pytorch-Kaldi \cite{Ravanelli2019} was then used to train the PPG extractor, which consists of 5 bidirectional GRU layers with 550 hidden units on each layer and a softmax output layer. In our experiments, we used 39-dim MFCCs plus delta features with a 25ms window and a 10ms shift. To reduce the dynamics, we took the logarithm scale on PPGs. Note that bilingual PPGs trained on two languages could also be used for better bilingual representability of PPGs \cite{zhou2019}.

\begin{figure}[t]
  \centering
  \includegraphics[width=7cm]{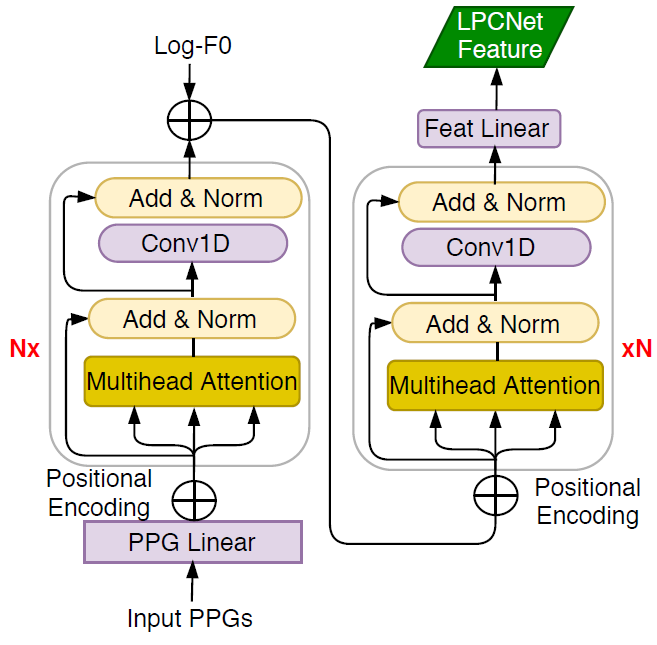}
  \caption{Illustration of the synthesizer network for FastSpeech-VC. The encoder is on the left and the decoder is on the right.}
  \label{fig:fastspeech_vc}
\end{figure}
\subsection{The synthesizer network}
We build the synthesizer network based on the FastSpeech TTS model \cite{Ren2019}, which belongs to a broad-class of encoder-decoder structure with self-attention mechanism. The FastSpeech uses only fully-connected (FC) networks with a non-autoregressive generation process. Both training and inference process can be parallelized, which significantly speeds up the conversion process. The FastSpeech model was originally designed to predict mel-spectrograms from text embedding. To perform VC tasks, we adapt FastSpeech model to predict LPCNet vocoder features from a sequence of PPGs and logarithm-scale fundamental frequencies(Log-F0) as shown in Fig \ref{fig:fastspeech_vc}. The text embedding is replaced by PPG embedding and Log-F0 is concatenated to the encoders' output.  The PPG embedding is generated by a fully connected (FC) linear layer (denoted as PPG Linear) which has 512 hidden layers and 512 output units. We also apply layer normalization and ReLU activation function to the outputs of PPG Linear.

In the synthesizer network, the encoder and the decoder are composed of a stack of N = 6 duplicate blocks. Each encoder block is composed of a multi-head self-attention and a 2-layer 1D convolutional neural network (CNN)
with residual connection and layer normalization. Triangular position encodings are added to the embedding before inputting to the encoder.
In our network, the duration predictor and the length regulator are removed from the original FastSpeech network, so that the input PPG sequence and the output LPCNet feature sequence have the same length. The output of the decoder feeds to a FC layer (denoted as Feat Linear) to predict the LPCNet feature.

The synthesizer network is a one-to-one mapping from a PPG frame to a LPCNet feature frame, which allows to adjust the converted speech rate by performing downsampling or upsampling of the PPG sequence during inference. In this respect, FastSpeech-VC is a controllable framework. Our synthesizer network was implemented based on the open-source ESPnet toolkit \cite{Shinji2018}.

\subsection{The LPCNet vocoder}
\label{sec:lpcnet_vocoder}
The LPCNet vocoder \cite{Valin2019} is a WaveRNN variant that combines linear prediction with recurrent neural networks to significantly improve the efficiency of speech synthesis. It reconstructs speech samples from the synthesizer's output, which consists of 18 Bark-scale cepstral coefficients and two pitch parameters (period, correlation). Compared to the 80-dim mel-spectrogram, the 20-dim LPCNet feature is more memory efficient. We use the open-source code published by Mozilla team on Github\footnote{https://github.com/mozilla/LPCNet} for the LPCNet vocoder.

\subsection{Training and conversion}
Similar to building a TTS system, we build FastSpeech-VC system in two stages: training stage, and inference or conversion stage.
At the training stage, all the three components: the PPG extractor, the synthesizer network, and the LPCNet vocoder, can be trained separately. The PPG extractor is speaker-independent and once trained can be used for any speech utterance. The synthesizer and LPCNet vocoder, on the other hand, are speaker-dependent and need to be trained for each target speaker. We first trained the PPG extractor
using AISHELL \cite{aishell2017} -- a large open-source Mandarin ASR corpus. Training the LPCNet vocoder is straightforward as we followed the instructions of the Github mentioned in section \ref{sec:lpcnet_vocoder}. To train the synthesizer, we first extracted three feature sequences: PPG, Log-F0 and LPCNet from the target speaker's corpus. The Log-F0 sequence was then normalized to be zero mean and unit variance. We used the training toolkit given in section 2.2 to train the synthesizer. A multi-speaker synthesizer is also possible as discussed in \cite{Ryan2018}. We however worked on single-speaker synthesizer and vocoder and leave the  multi-speaker synthesizer training as our feature work. Note that if there are only limited training data available for a target speaker, a pre-trained model should be employed.

At the conversion stage, MFCC and Log-F0 sequences were first extracted from the source speaker's utterances.
The PPG extractor was then used to transform the MFCC sequences to PPGs, which will be combined with the normalized Log-F0 sequences to feed into the target speaker's synthesizer to generate the LPCNet features. Finally, the target's LPCNet vocoder was used to generate the target's speech samples.

\section{Experiments}
\label{sec:experiments}
\subsection{Training setup}
We evaluated the proposed FastSpeech-VC on cross-lingual VC tasks and also included some results on mono-lingual VC tasks to show its generality. Three female target speakers were used: a native British English speaker (EN\_F), a native Mandarin speaker (CN\_F), and a native Japanese speaker (JP\_F). The British speaker has 27k utterances of English speech, the Mandarin speaker has 32k utterances of Mandarin speech, and for the Japanese speaker we used 500 utterances of Japanese speech. We trained FastSpeech-VC on the English and Mandarin target speaker separately, and adapted English FastSpeech-VC to the Japanese target speaker. We tested the target speakers' FastSpeech-VC using the other source speakers' speech. The source speakers include a  male native British English speaker (EN\_M) and a male native Mandarin speaker (CN\_M). The target speakers were also used source speakers of other target speakers. The speech of JP\_F were selected from JSUT TTS corpus \cite{jsut2017}. The other speakers' speech were selected from Alibaba TTS corpora. All speech were sampled at 16 kHz with 16-bit resolution.

\subsection{Baseline models}
We chose Tacotron2 \cite{Jonathan2018} and Transformer \cite{Naihan2018} TTS models as the baseline models to compare with the FastSpeech-VC in our proposed cross-lingual VC framework. We named the Tacotron2-based VC as Tracotron2-VC and the Transformer-based VC as Transformer-VC. The Tracotron2-VC was obtained by replacing the encoder and the decoder of the FastSpeech model by the encoder and the decoder of the Tacotron2 model \cite{Jonathan2018}. Except that we replaced the
character-embedding layer of the Tacotron2 TTS model with a FC linear layer for PPG embedding layer and changed the output layer to predict LPCNet feature, other components of the encoder and the decoder are retained as original. Our implementation for Tacotron2-VC was based on the Tacotron open-source code \footnote{https://github.com/keithito/tacotron}. In the experiments, the stop token loss weight was set to 1.0. The reduction factor was set to 3. The number mixture for the GMM attention was set to 8 and the attention kernel size was set to 7.

We similarly obtained the Transformer-VC by replacing the encoder and the decoder of the FastSpeech model with the encoder and the decoder of the Transformer  model \cite{Naihan2018}. Our implementation for Transformer-VC was closely following the recipe espnet/egs/ljspeech/tts1 of the open-source ESPnet toolkit [31].
Specifically, we used 8 attention heads and 1024 units for each
encoder and decoder block.  The stop token loss weight was 1.0 and the reduction factor was 1.
All synthesizer networks were optimized using Adam optimizer with a batch size of 32.

\begin{table}
\caption{The Naturalness MOS for the English target speaker with different cross-lingual VC systems. }
\begin{tabular}{lcccc}
\specialrule{.1em}{.05em}{.05em}
\multirow{3}{*}{Model} & \multicolumn{4}{c}{\multirow{2}{*}{Source Speaker}} \\
                       & \multicolumn{4}{c}{}                                           \\ \cline{2-5}
                       & EN\_M          & CN\_F         & CN\_M         & JP\_F         \\ \hline
GT                     & 4.41±0.11      & 4.22±0.13     & 4.14±0.12     & 3.70±0.14      \\ \hline
TA                     & 4.54±0.06      & 4.11±0.07     & 3.99±0.08     & 3.27±0.10     \\ \hline
TR                     & 4.47±0.07      & 4.16±0.07     & 3.99±0.08     & 3.32±0.11     \\ \hline
FS                     & 4.45±0.08      & 4.12±0.06     & 3.96±0.09     & 3.40±0.09      \\
\specialrule{.1em}{.05em}{.05em}
\end{tabular}
\end{table}

\begin{table}
\caption{The Naturalness MOS for the Mandarin target speaker with different cross-lingual VC systems. }
\begin{tabular}{lcccc}
\specialrule{.1em}{.05em}{.05em}
\multirow{3}{*}{Model} & \multicolumn{4}{c}{\multirow{2}{*}{Source Speaker}} \\
                       & \multicolumn{4}{c}{}                                           \\ \cline{2-5}
                       & EN\_F          & EN\_M         & CN\_M         & JP\_F         \\ \hline
GT                     & 4.45±0.15      & 4.38±0.14     & 3.99±0.15     & 3.51±0.16     \\ \hline
TA                     & 3.99±0.07      & 4.05±0.08     & 3.89±0.10     & 3.36±0.10     \\ \hline
TR                     & 4.01±0.08      & 3.98±0.09     & 3.89±0.09     & 3.42±0.09     \\ \hline
FS                     & 3.99±0.07      & 3.93±0.10     & 3.88±0.10     & 3.38±0.10     \\
\specialrule{.1em}{.05em}{.05em}
\end{tabular}
\end{table}
\subsection{Subjective test}
We conducted Mean Opinion Score (MOS) evaluations of speech naturalness and speaker similarity via subjective listening tests. Absolute Category Rating scale from 1 (bad) to 5 (excellent) was used with 0.5 point increments for naturalness and 0.5 point increments for similarity. For naturalness, raters were native speakers in the evaluating language. Each source-target speaker pair was assigned to a different group of raters. Each MOS score was averaged over 20 randomly picked utterances where each utterance was rated by at least 10 raters. For similarity, all the utterances were scored by the same group of raters, most of whom are (Mandarin-English) bilingual speakers. The raters were advised to ignore the spoken content for similarity tests. We showed the MOS scores with 95\% confidence intervals. Our denotations are: GT for ground-truth, TA for Tacotron2-VC, TR for Transformer-VC, and FS for FastSpeech-VC.

Table 1 and 2 showed MOS scores on speech naturalness. For English target speaker, FastSpeech-VC performs better than Tacotron2-VC and Transformer-VC on JP\_F and equivalently on other source speaker. For Mandarin target speaker, FastSpeech-VC performs as good as Tacotron2-VC and Transformer-VC. Compared to the ground-truth MOS, FastSpeech-VC has a gap ranging from 0.1 to 0.3 for the English target speaker and from 0.11 to 0.45 for the Mandarin target speaker, with similar gaps for mono-lingual VC results.

\begin{table}
\caption{The Similarity MOS for the English target speaker with different cross-lingual VC systems. }
\begin{tabular}{lcccc}
\specialrule{.1em}{.05em}{.05em}
\multirow{3}{*}{Model} & \multicolumn{4}{c}{\multirow{2}{*}{Source Speaker}} \\
                       & \multicolumn{4}{c}{}                                           \\ \cline{2-5}
                       & EN\_M          & CN\_F         & CN\_M         & JP\_F         \\ \hline
GT                     & \multicolumn{4}{c}{4.25±0.09}                                  \\ \hline
TA                     & 4.13±0.14      & 3.71±0.09     & 3.29±0.11     & 3.67±0.12     \\ \hline
TR                     & 4.00±0.14      & 3.67±0.11     & 3.33±0.09     & 3.79±0.11     \\ \hline
FS                     & 4.29±0.13      & 3.50±0.14     & 3.46±0.12     & 3.75±0.11     \\
\specialrule{.1em}{.05em}{.05em}
\end{tabular}
\end{table}

\begin{table}
\caption{The Similarity MOS for the Mandarin target speaker with different cross-lingual VC systems. }
\begin{tabular}{lcccc}
\specialrule{.1em}{.05em}{.05em}
\multirow{3}{*}{Model} & \multicolumn{4}{c}{\multirow{2}{*}{Source Speaker}} \\
                       & \multicolumn{4}{c}{}                                           \\ \cline{2-5}
                       & EN\_F          & EN\_M         & CN\_M         & JP\_F         \\ \hline
GT                     & \multicolumn{4}{c}{4.38±0.09}                                  \\ \hline
TA                     & 3.79±0.11      & 3.75±0.09     & 4.04±0.11     & 3.71±0.01     \\ \hline
TR                     & 3.79±0.09      & 3.71±0.07     & 3.83±0.11     & 3.79±0.11     \\ \hline
FS                     & 4.00±0.12      & 3.50±0.12     & 3.75±0.13     & 3.50±0.09     \\
\specialrule{.1em}{.05em}{.05em}
\end{tabular}
\end{table}
Table 3 and 4 showed MOS scores on speaker similarity. The ground-truth was evaluated using the original utterances of the speakers. The results showed that the FastSpeech-VC always achieved better scores for mono-lingual VC tasks and there were some variations in MOS scores for cross-lingual VC tasks. Compared to the Tacotron2-VC and Transformer-VC, the FastSpeech-VC has generally better scores for the English target speaker and slightly worse scores for the Mandarin target speaker. Compared to ground-truth, the FastSpeech-VC has gaps ranging from 0.5 to 0.88. The evaluations for the Japanese target speaker was not shown in our paper due to space limits, but we recommend the readers to refer to the audio samples from our demo page\footnote{Audio samples - https://alibabasglab.github.io/vc/}.

To show the controllability of the FastSpeech-VC, we obtained different speech rates ($\times 0.8$ and $\times 1.2$) by lengthening or shortening the PPG duration and attached the audio samples in our demo page.

We evaluated the inference latency of FastSpeech-VC compared with Tactron2-VC and Transformer-VC. Our results indicated that FastSpeech-VC speeds up the LPCNet feature generation more than 200 times and speeds up the overall audio generation more than 30 times.

\section{Conclusions}

We described a high-quality framework for cross-lingual voice conversion task based on neural FastSpeech TTS model and LPCNet vocoder. We proposed to use phonetic posteriorgrams and normalized logarithm-scale fundamental frequencies to overcome the significant phonetic and prosodic mismatches between two different languages.  Leveraging on the proposed effective feature representations and the advanced synthesizer and the vocoder, our proposed FastSpeech-VC produces high quality converted speech with good speaker similarity using mono-lingual corpus only. More importantly, FastSpeech-VC has the controllability to adjust speech rate smoothly and speeds up the audio generation significantly. In addition, FastSpeech-VC can easily be adapted to a speaker with only limited data.

\bibliographystyle{IEEEbib}
\bibliography{mybib}

\end{document}